Strain measurement in semiconductor FinFET devices using a novel moiré demodulation technique

V. Prabhakara[1,2], D. Jannis[1], A. Béché[1], H. Bender[2] and J. Verbeeck[1]

1. EMAT, University of Antwerp, Groenenborgerlaan 171, 2020, Antwerp, Belgium.

2. Imec, Kapeldreef 75, 3001, Leuven, Belgium.

**Abstract:**

Moiré fringes are used throughout a wide variety of applications in physics and engineering to bring out small variations in an underlying lattice by comparing with another reference lattice. This method was recently demonstrated in Scanning Transmission Electron Microscopy imaging to provide local strain measurement in crystals by comparing the crystal lattice with the scanning raster that then serves as the reference. The images obtained in this way contain a beating fringe pattern with a local period that represents the deviation of the lattice from the reference. In order to obtain the actual strain value, a region containing a full period of the fringe is required, which results in a compromise between strain sensitivity and spatial resolution. In this paper we propose an advanced setup making use of an optimised scanning pattern and a novel phase stepping demodulation scheme. We demonstrate the novel method on a series of 16 nm Si-Ge semiconductor FinFET devices in which strain plays a crucial role in modulating the charge carrier mobility. The obtained results are compared with both Nano-beam diffraction and the recently proposed Bessel beam diffraction technique. The setup provides a much improved spatial resolution over conventional moiré imaging in STEM while at the same time being fast and requiring no specialised diffraction camera as opposed to the diffraction techniques we compare to.

**Introduction:**

Strain has become an essential ingredient in the functioning and performance optimisation of modern semiconductor devices. As an example, tensile strain increases the electron mobility in n-MOS devices while compressive strain increases the hole mobility in p-MOS devices [1] [2] [3].With the ongoing trend of scaling down the size of transistors and the use of strain as a performance enhancing parameter, it is becoming essential to quantitatively measure strain at the nanoscale. Transmission electron microscopy based strain measurement methods provide the best possible spatial resolution, in principle down to the individual unit cell level. They can be broadly classified into diffraction and real space (imaging) based techniques. Diffraction based techniques like Nano-beam diffraction (NBD) offer a spatial resolution of 1 to 6 nm with a relative strain precision of $6 \times 10^{-4}$ [3]. The spatial resolution is limited in an attempt to keep the diffraction discs as small as possible in order to avoid complications due to redistribution of intensities within the discs stemming from multiple elastic scattering processes. Nano-beam electron diffraction with precession (N-PED) circumvents the problem by averaging out diffraction patterns acquired at different tilt angles, thus partially cancelling the influence of multiple scattering. The intensity in the diffraction spots becomes more uniform, improving the precision to better than $2 \times 10^{-4}$ [3]. Recently, Bessel beam diffraction was shown to provide similar performance as N-PED but with minimal hardware complexity [4]. All diffraction techniques have a very good precision but are usually slow due to hardware limited acquisition speed restrictions and long data analysis. Moreover, the probe size on the sample can vary in a sensitive way on defocus and lens aberrations, which can imply a change in spatial resolution between experiments.

Real space based techniques like High Resolution Scanning Transmission Electron imaging (HR-STEM) with GPA (Geometric phase analysis) have the ultimate spatial resolution as they are based on atomic resolution imaging of (a projection of) the crystal. In GPA, the spatial resolution and precision of the obtained strain map are inversely proportional and depend on the size of a mask applied in the Fourier space [5]. Typically, these methods have a lower field of view, but results are obtained fast due to rapid acquisition (a few seconds) and quick data analysis (less than a minute). STEM moiré strain mapping is an extension of this high resolution imaging which provides a higher field of view [6] at the expense of spatial resolution. The spatial resolution depends on the periodicity of the moiré fringes and is usually multiple times the crystal periodicity. In this paper, we attempt to extend the spatial resolution of the moiré strain analysis by combining it with phase shifting interferometry and demodulation techniques. We show that the precision obtained is approaching that of diffraction techniques while the spatial resolution can increase significantly and the acquisition and analysis time is lowered substantially.

**STEM moiré theory:**

STEM moiré is a phenomenon caused by interference between a chosen sampling grid and the crystal lattice. The sampling grid is the periodic arrangement of the points sequentially visited by the STEM probe. The illuminated crystal lattice is then sampled at these points. We assume the STEM probe to be infinitely small here, which allows us to express the scanning grid as a 2D Dirac comb.

The continuous image of the sample (independent of specific imaging mode in STEM) can be expressed as an intensity variation of a detector signal (e.g. a High Angle Annular Dark Field detector) depending on the probe position $I(y, z)$.

Sampling in STEM is analogous to multiplication of this image function by a 2D Dirac delta comb $S(y, z)$

$$S(y,z) = \sum_{m=-\infty}^{m=\infty} \sum_{n=-\infty}^{n=\infty} \delta(y - mp_y)\delta(z - np_z) \qquad (1)$$

With $p_y$ and $p_z$ the spatial sampling step in y and z directions. We get a resulting discrete image:

$$I(y_m, z_n) = I(mp_y, np_z) \qquad (2)$$

In frequency space, this becomes a convolution of the Fourier transform (FT) of the image $I(u, v)$ with the FT of the 2D sampling grid which transforms to:

$$S(u,v) = \sum_{m=-\infty}^{m=\infty} \sum_{n=-\infty}^{n=\infty} \delta(u - mf_{py})\delta(v - nf_{pz}), \qquad (3)$$

with, $f_{py} = \frac{1}{p_y}, f_{pz} = \frac{1}{p_z}$.

The discrete frequency spectrum of the sampled image is then written as

$$I_c(u,v) = I(u,v) * \sum_{m=-\infty}^{m=+\infty} \sum_{n=-\infty}^{n=+\infty} \delta(u - mf_{py})\delta(v - nf_{pz}) \qquad (4)$$

Note that we have taken the finite size of the sampling grid (the field of view) as a cut-off into the image spectrum rather than in the Fourier transform of the grid to keep the delta functions.

$$I_c(u,v) = \sum_{m=-\infty}^{m=+\infty} \sum_{n=-\infty}^{n=+\infty} I(u - mf_{py}, v - nf_{pz}) \qquad (5)$$

This results in an infinite series of replicas with the central one occupying the frequency range $\left[\frac{-1}{2p_y}, \frac{1}{2p_y}\right] \cup \left[\frac{-1}{2p_z}, \frac{1}{2p_z}\right]$. If the maximum frequency in the image $u\_\max < 1/2\, p_y$ and $v\_\max < 1/2\, p_z$, the sampling satisfies the Nyquist criterion and each replica is separated from all other replicas(Figure.1b). In the case of under-sampling, aliasing happens, leading to the formation of moiré fringes by the replication of frequencies from outside the central portion into this region (Figure.1c).

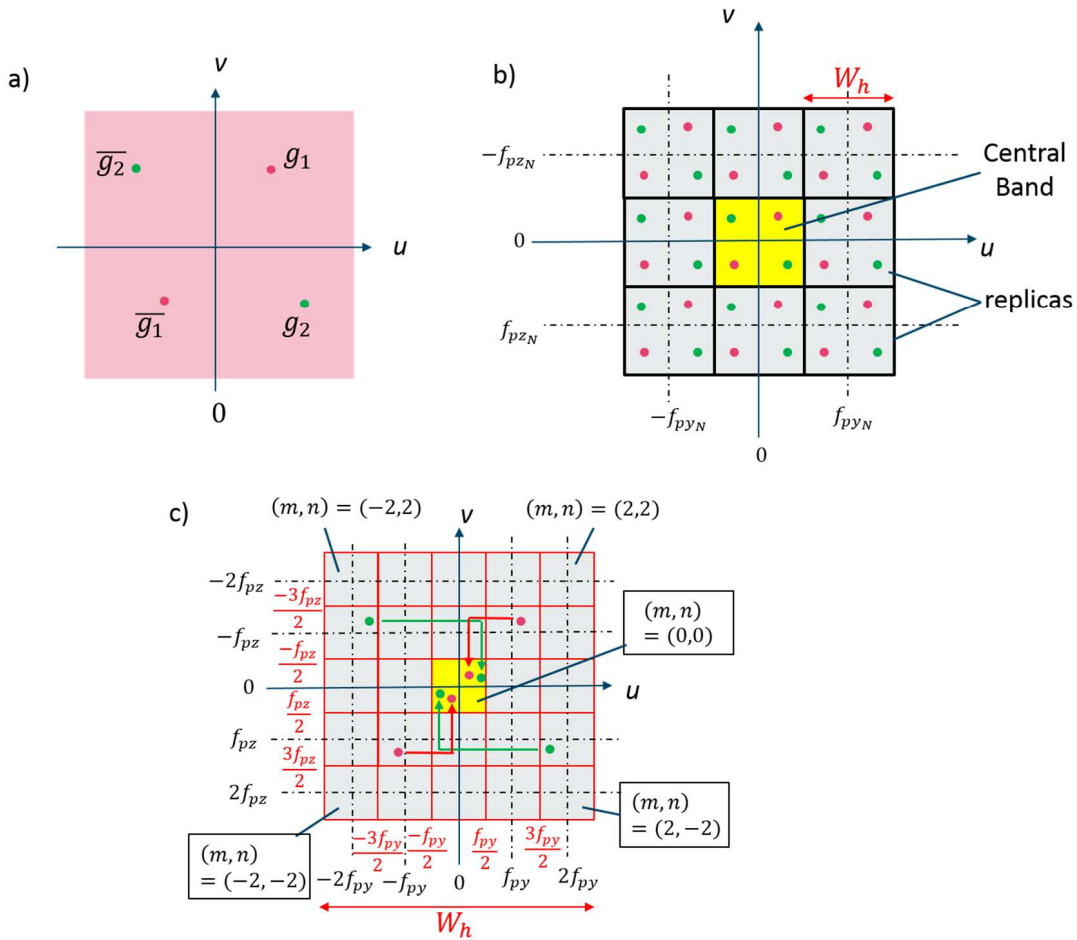

*Figure 1 a) Frequency represenation of a periodic 2D signal. b) Sampling with Nyquist criterion results in an infinite amount of replicas indicated by grey boxes with dimensions determined by the sampling frequency. c) moiré effect with sampling frequency lower than the highest frequency. The replication due to the finite sampling*

*results now in a mixing up of the position of the frequency components which is best noted in the central replica marked in yellow. This mixing up is the essence of the moiré effect where frequencies appear different from what they really were without sampling. The frequencies get aliased and a high resolution frequency spectrum is now divided into smaller moiré windows with index (m,n) due to a lower sampling frequency in the moiré compared to a high resolution image. The arrow indicates the translation of frequencies into the new smaller central band.*

As an example, we can assume an image made of two sinusoids ($g_1, g_2$) as shown in Figure 1a. A sampling frequency $f_{py}$ and $f_{pz}$ is chosen such that the Nyquist criterion is not satisfied, which results in the moiré effect (Figure 1c). The yellow window represents the central replica with each red box an identical replication. The index of each window is represented by (m,n) and the frequencies that are outside the yellow window are translated into the yellow window by shifting the window over ($mf_{py}, nf_{pz}$). For example, the frequency g₁ is shifted one window down in v and one window left in u and so on [7]. This leads to a new apparent spatial frequency (the moiré frequency) within the central frequency band:

$$g'_y = g_y - mf_{py} \qquad (6)$$

$$g'_z = g_z - nf_{pz} \qquad (7)$$

Where $g_y$ and $g_z$ are the frequency components in y and z directions. For a projection image of Si or Ge along the [110] zone axis, we consider the y axis as the [1$\bar{1}$0] direction and the z axis along the [001] direction. The lattice distances for Si are different in these two directions $a_{1\bar{1}0} = 0.384\ nm$ and $a_{001} = 0.543\ nm$ (Figure 2a). This means, for any fundamental frequency in y and z directions, $g_y$ and $g_z$, we have $g_y > g_z \in (u,v)$. In conventional STEM, a square scanning grid is used and $f_{py} = f_{pz}$. In general, this results in the creation of 1D moiré as m and n are not necessarily the same.

Figure 2d shows an experimental image obtained with a square scanning grid, resulting in only one fringe direction to be visible while the other is too high in frequency. These 1D moiré fringes are easily observed for a probe aberration corrected instrument (the probe has to be smaller than the lattice spacing) and readily reveal strain as a slight change of the moiré fringes in different locations of the image.

The resulting distortion of the apparent symmetry by the formation of 1D moiré can be circumvented and a more interpretable situation occurs if the symmetry of the scanning grid matches the symmetry of the lattice by adjusting the scan pattern to obey:

$$\frac{g_y}{f_{py}} = \frac{g_z}{f_{pz}} = U, \qquad (8)$$

with $U$ the under-sampling factor > 0.5. This can be done by adjusting the strength of the signals driving the scan coils. In this case, all frequencies transform as:

$$g'_y = g_y - mf_{pz}\frac{g_y}{g_z} = g_y\left(1 - m\frac{f_{pz}}{g_z}\right) = g_y\left(1 - \frac{m}{U}\right) \qquad (9)$$

$$g'_z = g_z - nf_{py}\frac{g_z}{g_y} = g_z\left(1 - n\frac{f_{py}}{g_y}\right) = g_z\left(1 - \frac{n}{U}\right) \qquad (10)$$

If m=n, this transform is an angle preserving scaling factor and the moiré fringes will have the same symmetry as the lattice. An experimental example is shown in Figure.3a.

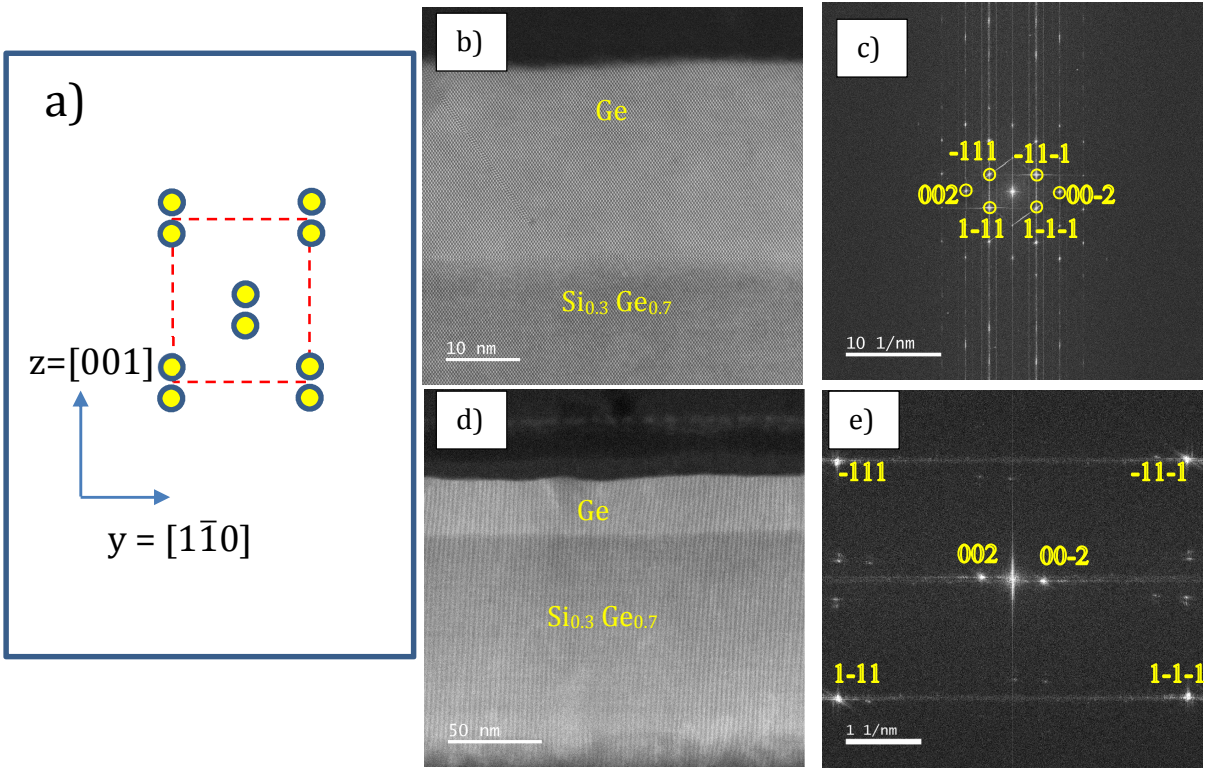

Figure 2 a) Diagram of the unit cell of Si and Ge in [110] zone axis b) HRSTEM image of the Si and Ge material and its corresponding diffractogram c). d) 1D moire created on the Si and Ge material by undersampling at lower magnification and its corresponding diffractogram e). The undersampling results in clear moire fringes in d) caused by only the {002} spots being replicated to rather low frequencies and a mixing up of the frequency components in e). The moiré fringes are sentive to strain as seen by the slight wavyness and tilting of these fringes as a function of the location in the sample.

When strain occurs in the lattice, frequencies will shift by $\Delta g$ and a relative shift of the diffracted spots can be obtained in the diffraction pattern (as in NBD). The strain in a particular direction is expressed as:

$$\varepsilon = \frac{|\Delta \vec{g}|}{|\vec{g}|} \quad (11)$$

When under-sampling occurs, an apparent strain will appear in the moiré fringe as

$$\varepsilon' = \frac{|\Delta \vec{g}|}{|\vec{g'}|} = \frac{|\Delta \vec{g}|}{\sqrt{(g_x - mf_{px})^2 + (g_y - nf_{py})^2}} = \alpha \varepsilon \quad (12)$$

The strain gets magnified by a boost factor $\alpha$ which makes it much more apparent as can be seen both in Figure 3a and Figure 4. This boosting is shape preserving ($\varepsilon'$ stays parallel to $\varepsilon$) only in the case where the scan grid matches the symmetry of the crystal and a situation is chosen where m=n. The strain boosting factor is plotted as a function of the under-sampling factor U showing how strain boosting is especially strong for low frequency moiré fringes. This also implies a hard trade-off between strain sensitivity and spatial resolution. In order to determine the frequency of the moiré fringe (related to the strain), one needs at least an area of the order of a period of the fringe.

Another limitation comes from the fact that strain should not shift frequencies outside the central frequency replica. Otherwise m and n may differ for different areas of the sample leading to a much more complicated analysis (in fact a multivalued problem with non-unique solution). This limits the tolerable strain variation in a single image as strain amplification goes up.

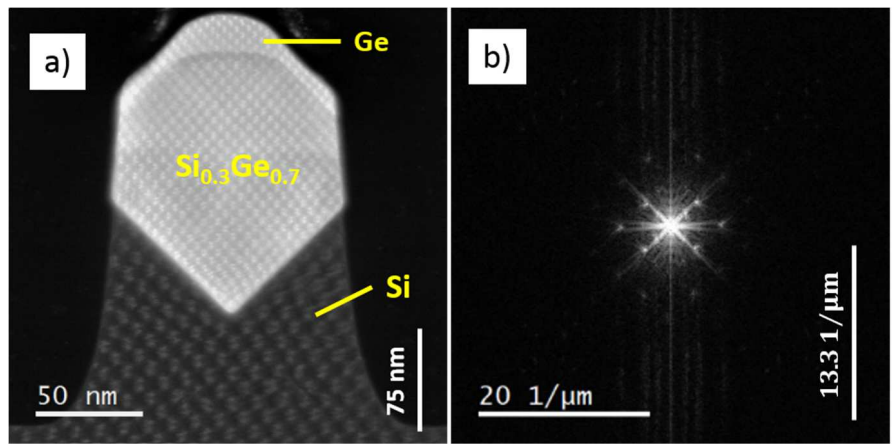

Figure 3a) 2D moiré fringes on a 100nm FinFET cross section in [110] zone axis and its corresponding b) diffractogram. Note the effect of local strain becoming visible as a distortion of the moire fringe pattern in different locations of the sample.

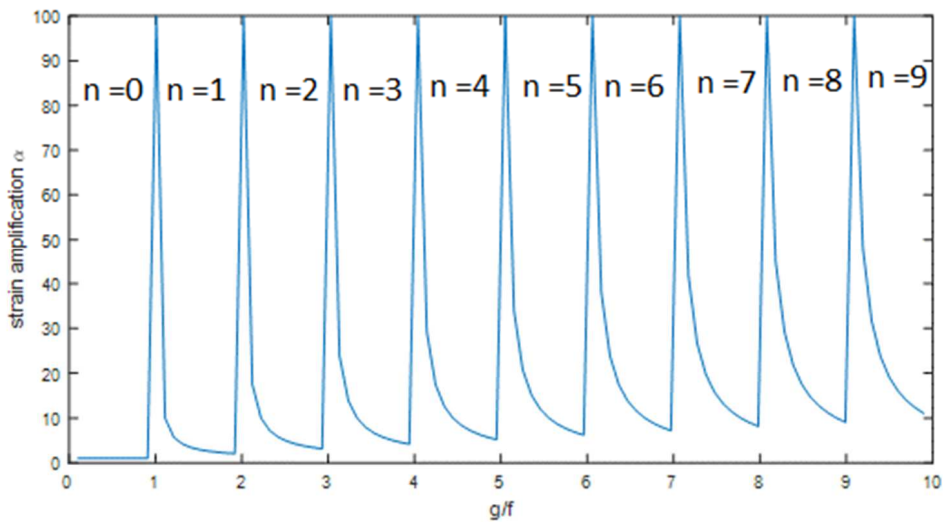

Figure 4 Strain amplification factor as a function of under-sampling. Note how strain magnification depends in a sensitive and nonlinear way on the choice of $U = g/f$ and n is the moiré window(See Appendix A). The value of n equal to zero is for an oversampled high resolution image and it does not have any amplification ($\alpha$=1). The strain amplification is extremely high when n = U.

**Recovering the strain from Moiré images: conventional way**

Strain analysis of moiré STEM images is typically performed using Geometric phase analysis (GPA) [8]. In this method a mask is used in the frequency domain to isolate the side bands around selected frequencies that carry the strain information. This selection mask determines also the spatial resolution that can be obtained and determines the signal to noise ratio in the resulting strain map. The larger the radius of the mask, the higher is the spatial resolution and the more noise enters the estimated strain map [8]. The mask radius is chosen to prevent selecting more than one diffraction spot at a time and is typically chosen smaller than half of the selected reference frequency. This reference lattice vector is used to demodulate the measured signal which then reveals the strain with respect to this reference.

Applying this method to moiré STEM images reveals the apparent strain which needs to be corrected back to the actual strain via eq.[12] as shown in fig.5. As the strain magnification is significantly higher for low frequency moiré fringes, it is tempting to choose the undersampling as demonstrated in fig.5.a where strain is readily apparent from the moiré image due to the boosting effect. The low frequency moiré fringes will however demand a small selection mask and result in poor spatial resolution. On top of this, multiple spatial frequencies, possibly having different m and n values, tend to cluster near the centre and can be very close to each other. This results in unwanted overlaps and the strain maps become unreliable, especially as each of these frequencies should be converted from apparent strain to actual strain with different scaling factors (Figure 5).

In order to overcome this difficulty and increase the practical usefulness of moiré strain mapping, we developed an alternative method that allows significantly suppressing other unwanted replica's.

**Extending the spatial resolution with quadrature demodulation**

If we could suppress the unwanted frequencies in the undersampled diffractogram, we could use a much larger mask size before demodulating the signal back to an image of the strain. This would result in higher spatial resolution and the suppression of artefacts. This is achieved by moiré phase stepping interferometry and Quadrature demodulation. The principle of moiré phase stepping interferometry is to phase shift the moiré fringes in different specific directions in repeated experiments.

If we shift the experimental scanning grid over $\Delta y$ and $\Delta z$ in y and z directions,

$$S(y, z) \rightarrow S(y - \Delta y, z - \Delta z) \qquad (13)$$

and apply the Fourier shift theorem we get for the Fourier transform of the sampling grid

$$S(u, v) \rightarrow S(u, v) e^{i2\pi(u\Delta y + v\Delta z)} \qquad (14)$$

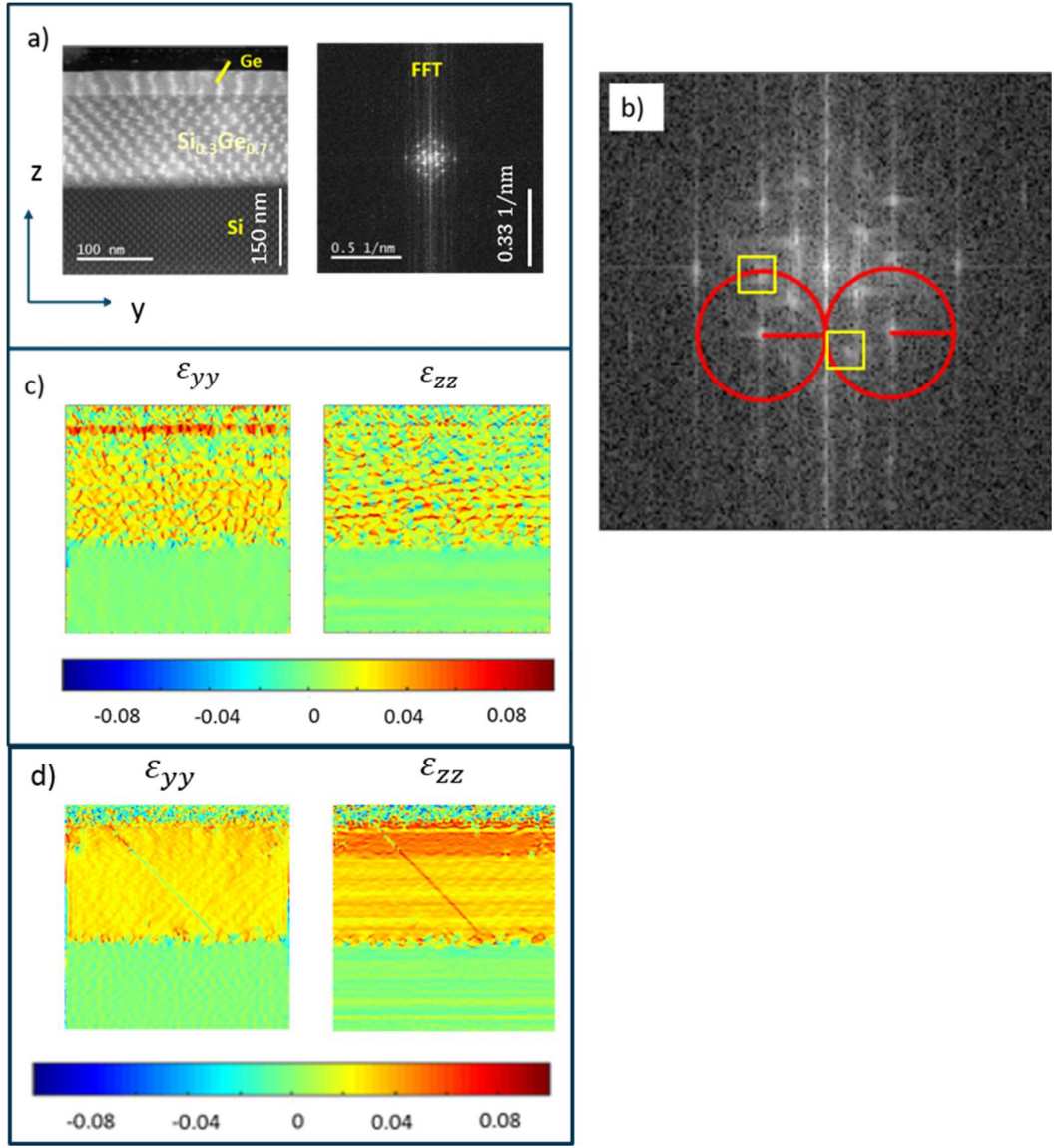

*Figure 5a) 2D moiré on the 16 nm Fin long section and its corresponding FFT b) GPA Masks used to isolate the regions in the FT. The yellow squares are the undesired interfering frequencies c) Normal strain $\varepsilon_{yy}$ and $\varepsilon_{zz}$ in the y and z directions obtained by GPA d) Comparing normal strain $\varepsilon_{yy}$ and $\varepsilon_{zz}$ in the y and z directions obtained by GPA(using the same size of the mask) on moiré with quadtrature demodulation. Since the size of the mask used here is small, the spatial resolution is also reduced. A reliable strain map is obtained with the application of Quadrature demodulation.*

This results in phase shifting of the different components in the spectrum as

$$I(u,v) = \sum_{m=-\infty}^{m=\infty} \sum_{n=-\infty}^{n=\infty} I(u - mf_{my}, v - nf_{mz})e^{i2\pi(u\Delta y + v\Delta z)} \quad (15)$$

This equation shows that it is sufficient to shift the sampling grid by a reference lattice fringe spacing $\Delta y = \frac{1}{g_y}$ and $\Delta z = \frac{1}{g_z}$ to cause a $2\pi$ phase shift of these moiré fringes in the under-sampled image.

Now, that we have the capability to phase shift the frequency components in the recorded moiré images by spatially shifting the sampling grid, we can apply an extended version of quadrature demodulation. Quadrature demodulation is commonly used in telecommunication (e.g. FM radio) [9] and allows to extract a given sideband of a modulated signal. In our case the 'carrier frequency' is a reference spatial frequency and the 'signal' is the strain or deviation with respect to this signal.

In this case we apply a 4x4 array of shifts of the scanning pattern, resulting in 16 independent and phase shifted moiré images. The applied shift can be represented as a 4x4 matrix S targeted to bring out the strain along 2 non-collinear reference frequencies $g_1$ and $g_2$.

$$\vec{S} = \begin{bmatrix} (0,0) & \left(\frac{1}{8g_{y1}}, \frac{1}{8g_{z1}}\right) & \left(\frac{2}{8g_{y1}}, \frac{2}{8g_{z1}}\right) & \left(\frac{3}{8g_{y1}}, \frac{3}{8g_{z1}}\right) \\ \left(\frac{1}{8g_{y2}}, \frac{1}{8g_{z2}}\right) & \left(\frac{1}{8g_{y1}}+\frac{1}{8g_{y2}}, \frac{1}{8g_{z1}}+\frac{1}{8g_{z2}}\right) & \left(\frac{2}{8g_{y1}}+\frac{1}{8g_{y2}}, \frac{2}{8g_{z1}}+\frac{1}{8g_{z2}}\right) & \left(\frac{3}{8g_{y1}}+\frac{1}{8g_{y2}}, \frac{3}{8g_{z1}}+\frac{1}{8g_{z2}}\right) \\ \left(\frac{2}{8g_{y2}}, \frac{2}{8g_{z2}}\right) & \left(\frac{1}{8g_{y1}}+\frac{2}{8g_{y2}}, \frac{1}{8g_{z1}}+\frac{2}{8g_{z2}}\right) & \left(\frac{2}{8g_{y1}}+\frac{2}{8g_{y2}}, \frac{2}{8g_{z1}}+\frac{2}{8g_{z2}}\right) & \left(\frac{3}{8g_{y1}}+\frac{2}{8g_{y2}}, \frac{3}{8g_{z1}}+\frac{2}{8g_{z2}}\right) \\ \left(\frac{3}{8g_{y2}}, \frac{3}{8g_{z2}}\right) & \left(\frac{1}{8g_{y1}}+\frac{3}{8g_{y2}}, \frac{1}{8g_{z1}}+\frac{3}{8g_{z2}}\right) & \left(\frac{2}{8g_{y1}}+\frac{3}{8g_{y2}}, \frac{2}{8g_{z1}}+\frac{3}{8g_{z2}}\right) & \left(\frac{3}{8g_{y1}}+\frac{3}{8g_{y2}}, \frac{3}{8g_{z1}}+\frac{3}{8g_{z2}}\right) \end{bmatrix} \quad (16)$$

Each index of the matrix can be written as:

$$\vec{S}_{k,l} = \frac{k}{8\vec{g}_1} + \frac{l}{8\vec{g}_2}. \quad (17)$$

As a spatial shift of $1/g$ implies a $2\pi$ phase shift, the shifts result in a complex factor that is applied to a specific frequency component in the frequency domain:

$$f_{kl}(\vec{g}) = \exp(i2\pi \vec{g} \cdot \vec{S}_{kl}) \quad (18)$$

This means that we can single out any specific spatial frequency of interest by multiplying each sub-diffractogram $I(u,v)_{kl}$ with the complex conjugate of the above matrix and then summing up over all k,l indices. This results in constructive interference of a selected frequency g while destructive interference occurs approximately for all other frequencies

$$I_{demod}(u,v)_{g_y,g_z} = \sum_{kl} I(u,v)_{kl} f_{kl}^*(g_y,g_z)) * \delta(u - g_y, v - g_z) \quad (19)$$

with the delta function doing the actual demodulation. Converting this back to real space leads to the demodulated strain component with respect to the selected reference frequency.

$$I_{demod}(y,z)_{g_y,g_z} = \mathcal{F}^{-1}\left(I_{demod}(u,v)_{g_y,g_z}\right) \quad (20)$$

The selectivity improves with a $n \times n$ shifting matrix with increase in $n$, but this comes at the expense of an increase in the measurement time as $n^2$. A minimal choice would be n=2, but we show in the experimental section that n=4 provides a more realistic compromise. In fact, the suppression rate for any other frequency g in the discrete Fourier space of the image can be calculated as.

$$SR(g, g_{desired}) = \frac{1}{16}\sum_{k,l} f_{k,l}(g) f^*{}_{k,l}(g_{desired}) < 1 \qquad (21)$$

Ideally this suppression preserves frequencies close to the desired frequency as these carry the strain information but achieves a high suppression of frequencies that are further out.

**Experiment:**

In order to implement the above mentioned idea in the transmission electron microscope, it is necessary to obtain software control over the probe scan engine. Making use of a custom built scan engine[collaboration M. Tence and M. Kociak [10]], we are able to freely program the scanning pattern in an aberration corrected FEI Titan³ STEM instrument operating at 300 kV.

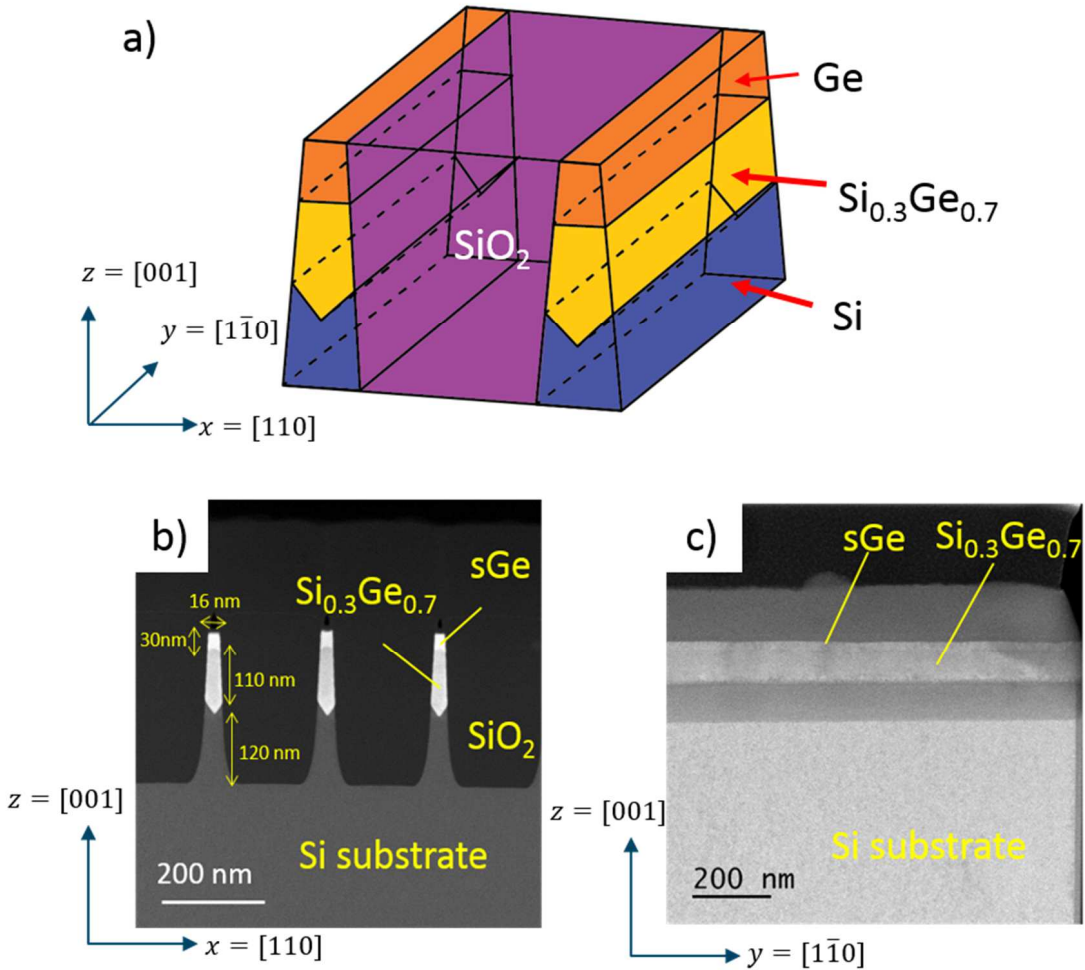

*Figure 6 Schematic description of the FinFET b) Cross section and c) long section of the 16 nm FinFET for the TEM strain analysis.*

We start with a 16 nm Finfet test sample cut along the longitudinal section indicated in Figure.6 to demonstrate the method. This sample consists of a silicon substrate on top of which a thick Si$_{0.3}$Ge$_{0.7}$ layer(~110nm) is grown epitaxially which is relaxed completely (due to a high aspect ratio, the structure relaxes by the formation of dislocations) and acts as the strain relaxed buffer (SRB). Then a thin layer(~30nm) of germanium is grown on top of SRB which is the channel of the FinFET. The Fins are 16nm in width and are separated by silicon oxide (Shallow Trench Isolation, STI). The germanium channel FinFET material is developed by IMEC using the STI-first technique [11]. The cross section and long section of the Fins for the TEM analysis under low magnification STEM mode is shown in the Figures 6b and 6c. The V shape of the silicon substrate in the cross section is used especially to trap the defects in the SRB and prevent their propagation into the Ge channel [11].

The scanning step-size is tuned to 17/16$^{th}$ of the lattice spacing of the silicon substrate in the y and z direction, i.e., $p_{yh} = \frac{17}{16} d(Si)_{1\bar{1}0}$ and $p_{zh} = \frac{17}{16} d(Si)_{001}$. This results in a distorted under-sampled (U=17/16=1.0625) moiré image that images the rectangular [110] projection of the silicon and germanium lattice as a square. This gives n=m=1 and brings us into the central band with a strain amplification factor of $\propto = 17$ (eq.12).

For each scan point, 16 slightly shifted probe positions are visited with a shift vector given by eq. 17 with $g_1 = \frac{1}{d_{1\bar{1}0}}$ and $g_2 = \frac{1}{d_{001}}$. This results in a shift of the sampling probe by

$$\vec{S}_{k,l} = k\Delta y \hat{y} + l\Delta z \hat{z} \ (with \ k, l = 0 \dots 3) \tag{22}$$

with $\Delta y$ and $\Delta z$ chosen to be $\Delta y = \frac{d_{1\bar{1}0}}{8}$ and $\Delta z = \frac{d_{001}}{8}$ as sketched in Figure 7.

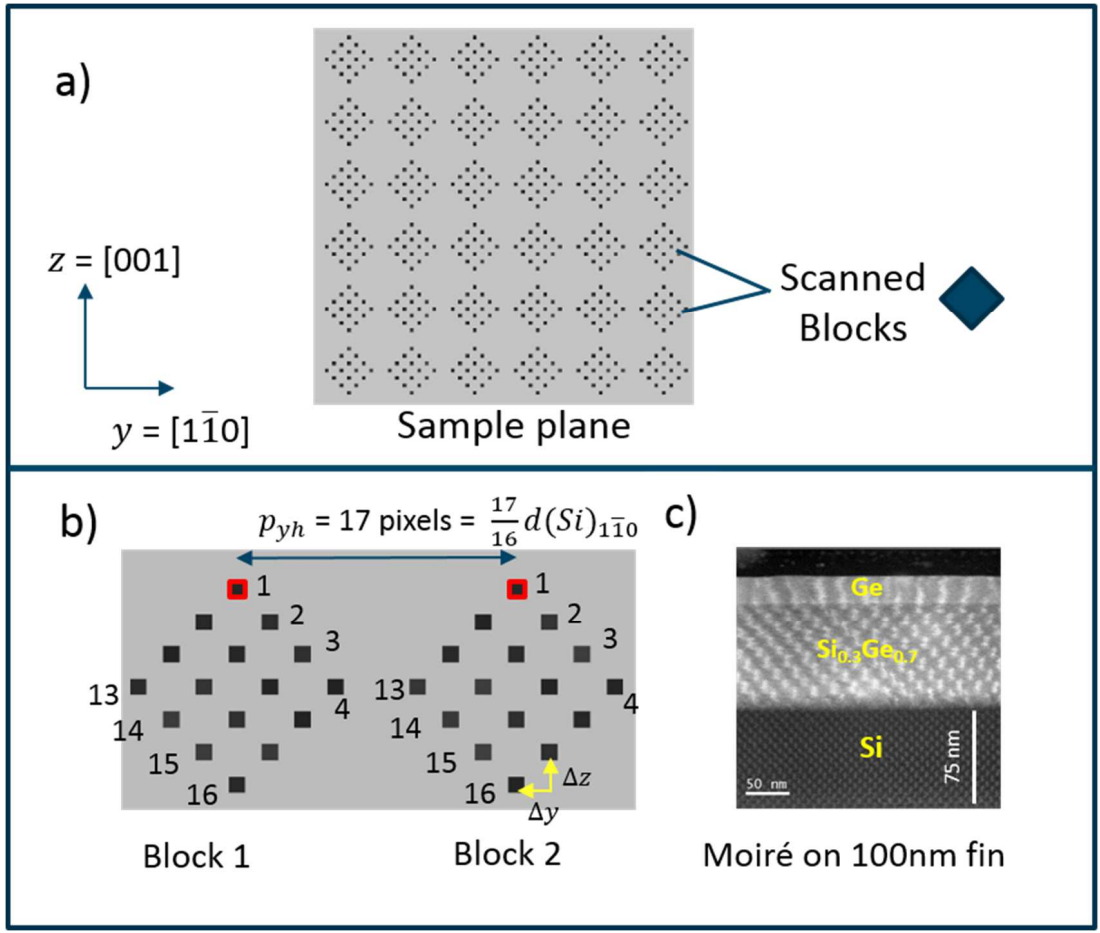

*Figure 7a) Block scanning pattern overview. Black squares represent probe positions that are visited sequentially in blocks of 4x4 pixels. The order of the probe visiting positions inside each block is shown in fig.7b. c) moiré image obtained by using only pixel 1 from each block, marked in red in b. The pixel step size is $d_{1\bar{1}0}/16$ in y direction and $d_{001}/16$ in z-direction. This makes the pixel size for the moiré image $^{17}/_{16}$ th of the lattice spacing of Silicon($d_{1\bar{1}0}$ or $d_{001}$ depending on the direction).*

This results in 16 independent moiré images with shifted fringes. Applying the quadrature demodulation as explained above, we obtain the demodulated result for $g = g_0, g = g_1, g = g_2$ as shown in Figure 8.

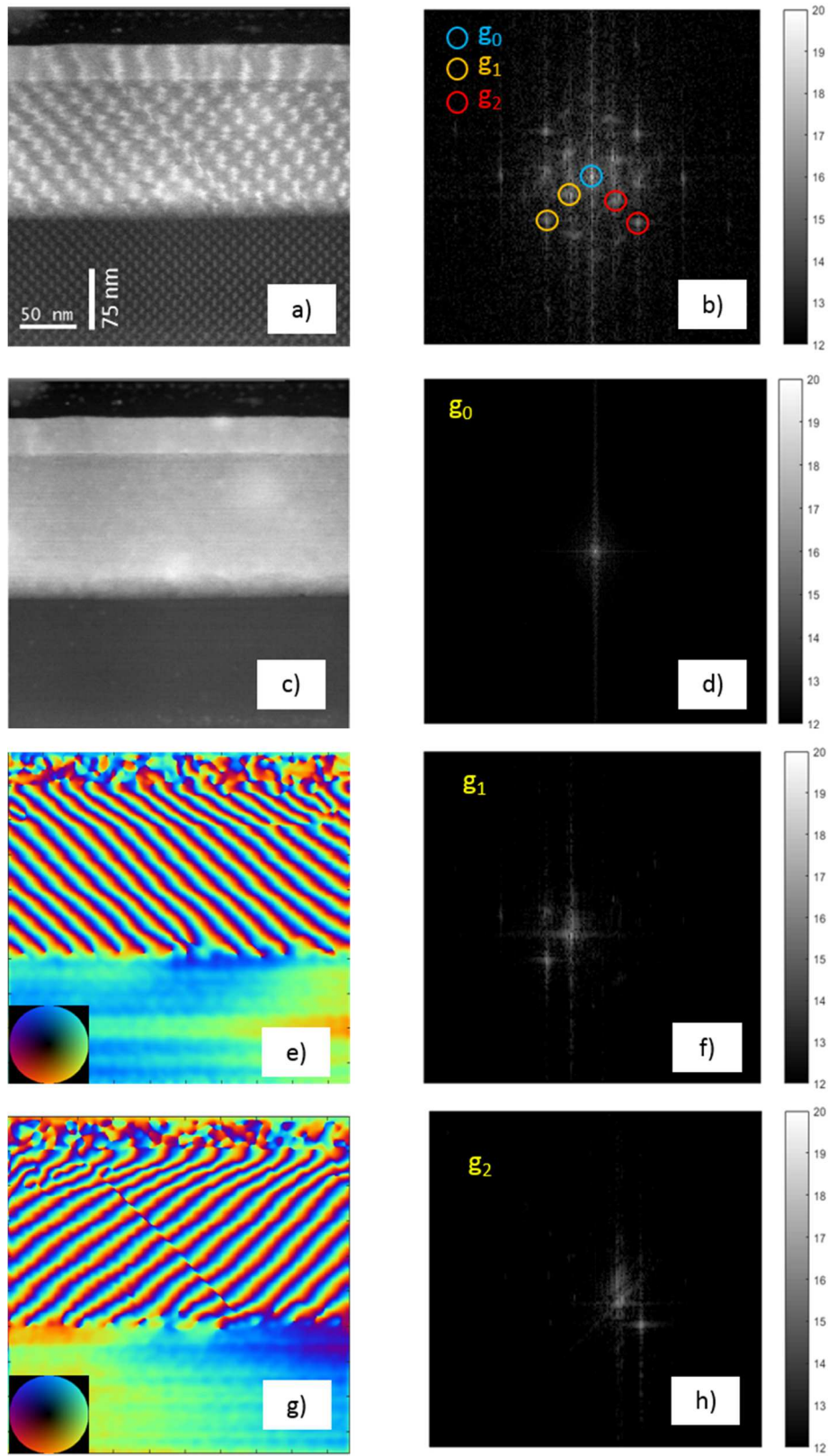

*Figure 8 a) Moiré on the 16nm Fin longitudinal direction b) and its diffractogram c) Demodulation to either g1 and g2 with a 4x4 shift matrix is shown as the phase of the demodulated wave in y,z-space (e,g) and g0 with the real value in x,y-space (c) and the absolute value of amplitude in frequency space of $g_0$, $g_1$ and $g_2$ (d,f,h). Note the selection of the desired frequencies while a strong suppression of the unwanted frequencies occur. There are two reflections in g1 and g2 corresponding to Si and $Si_{0.3}Ge_{0.7}$ material having different lattice spacings, but both are maintained after the quadrature demodulation process selecting Si as the reference lattice. This shows that the strained components remain while the other unwanted components are strongly suppressed.*

We can calculate the suppression factor for each of the selected frequencies $g_0$, $g_1$, $g_2$ (Figure 9). Quadrature demodulation keeps the amplitude of the selected frequency while suppressing the other unwanted frequencies by destructive interference. The suppression factor of the undesired frequency is the damping factor that is applied to the undesired frequencies while maintaining the desired frequencies. Figure.9 shows that the effect of quadrature demodulation is similar to applying a 2D sinc function enhancement factor centred at the selected frequency while the other frequencies away from the selected frequency are greatly suppressed. As an example, if the frequency $g_1$ is selected, the suppression factor for the other two frequencies $\frac{Ig_0}{Ig_1} = 4.5e^{-17}$ and $\frac{Ig_2}{Ig_1} = 1.5e^{-16}$ i.e., the ratio of the absolute value of amplitude of undesired frequency to that of the selected frequency in the Fourier space.

This suppression of the undesired frequencies is highly beneficial when demodulating the reference frequency to zero as it allows the use of a very relaxed low pass filter (possibility to cover the full Fourier space of the image) to supress only far away components resulting in a much improved spatial resolution up to 1 nm.

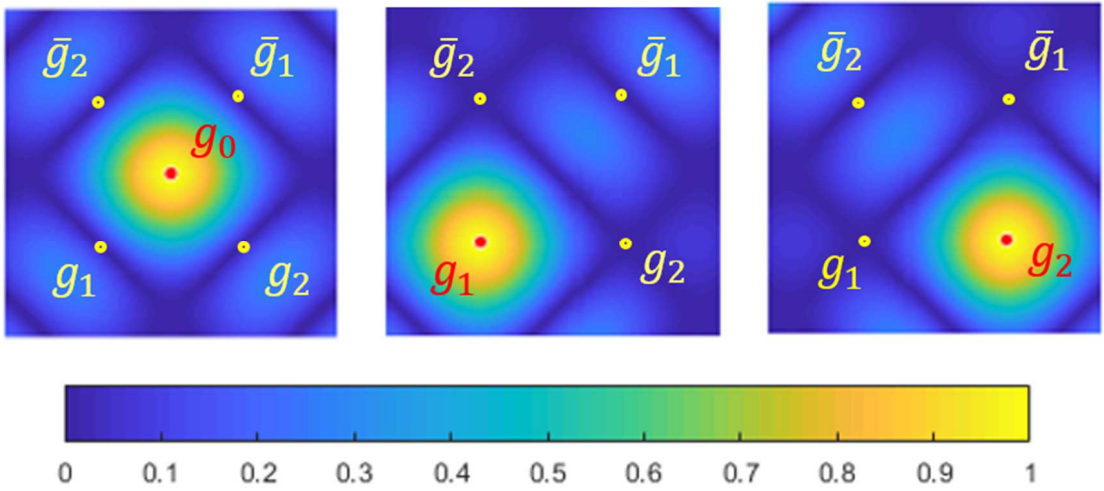

*Figure 9 Quadrature demodulation scheme results in the extraction of the desired frequency bands. The figure illustrates the enhancement factor applied on three desired frequency sidebands $g_0$, $g_1$ or $g_2$. The centres desired frequencies $g_0$, $g_1$ and $g_2$ are highlighted with the red dot and the centres of undesired frequencies are highlighted in yellow.*

To extract the 2D strain tensors we need to demodulate and extract strain with respect to two non-coplanar reference frequencies $g_1$ and $g_2$. The demodulated phase components $Pg_1$ and $Pg_2$ along both diagonal directions are shown as a phase plot indicating phase as colour (Figure 8e and 8g) and amplitude as intensity in Figure 8c.

Strain is extracted using the formula derived from Hÿtch et al. [8]. The displacement field is given as:

$$\begin{pmatrix} u_y \\ u_z \end{pmatrix} = -\frac{1}{2\pi} \begin{pmatrix} g_{1y} & g_{1z} \\ g_{2y} & g_{2z} \end{pmatrix}^{-1} \begin{pmatrix} P_{g1} \\ P_{g2} \end{pmatrix} \quad (23)$$

Where, $u_y$ and $u_z$ are the y and z components of the displacement field $u$. $g_{1y}$ and $g_{2y}$, $g_{1z}$ and $g_{2z}$ are the y and z components of the two non-coplanar side bands of interest.

The distortion tensor $\mathcal{D}$, strain tensor ε and rotation tensor ω can be calculated from the gradient of the displacement field,

$$\mathcal{D} = \begin{pmatrix} \partial u_y/\partial y & \partial u_y/\partial z \\ \partial u_z/\partial y & \partial u_z/\partial z \end{pmatrix} = \begin{pmatrix} \mathcal{D}_{yy} & \mathcal{D}_{yz} \\ \mathcal{D}_{zy} & \mathcal{D}_{zz} \end{pmatrix} \quad (24)$$

$$\varepsilon = \frac{1}{2}(\mathcal{D} + \mathcal{D}^T) \quad (25)$$

$$\omega = \frac{1}{2}(e - e^T), \text{local rigid rotation} \quad (26)$$

**Comparison to alternative techniques**:

In order to appreciate the performance of the present method, we compared it to the diffraction based strain measurement techniques like Nano-beam diffraction(NBD) and Bessel diffraction [12] [4] [3]. The analysis with these methods is applied to a 16nm FinFet, which is a challenging structure in terms of the spatial resolution requirement but a realistic example of the current semi-conductor technology.

NBD was performed using a quasi-parallel probe with a beam convergence angle of ≈ 0.3 mrad at 300kV using an FEI TITAN³ microscope. Bessel diffraction was performed at 300kV using an aberration corrected FEI TITAN microscope and using an annular aperture in the C2 aperture plane. The convergence angle was chosen to be ≈6 mrad and this gives a diffraction pattern consisting of rings instead of discs (Figure.10). Such lower convergence angle is chosen so as to limit the overlapping area between rings. This improves the precision of the strain measurement but is traded off with the spatial resolution [4].

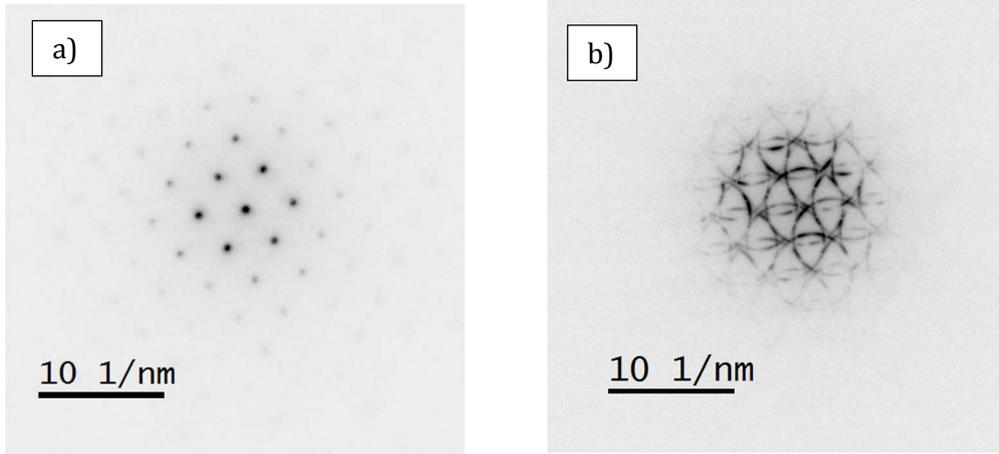

*Figure 10 Diffraction pattern obtained with the NBD a) and Bessel b) diffraction techniques on Si at [110] zone axis. A spot diffraction pattern is obtained in NBD due to the quasi-parallel probe, while ring diffraction patterns are obtained in Bessel diffraction due to the annular aperture and a convergent probe.*

The NBD data is processed with the Epsilon 1.2.1.40 software from Thermofischer scientific. The Bessel data is treated with a dedicated python based software, which is based on the autocorrelation of each diffraction pattern and subsequent extraction of the centre of the rings by background fitting and normalization [4].

The strain maps were also obtained with the discussed phase stepping method, and Quadrature demodulation offering an estimated spatial resolution of 2.8 nm (Figure 12). Line profiles across the strain maps are averaged over 16 nm width perpendicular to the scan direction. Some scanning artefacts in the slow scan direction are still visible in e.g. the $\epsilon_{zz}$ strain line profile as stripe artefacts in the horizontal direction. These kinds of artefacts are typical for the STEM based strain methods and can be due to sample drift [13] and electromagnetic noise which results in instabilities of the probe position with respect to time. The precision of the proposed technique in the fast scan direction is estimated as 1.1 x 10$^{-3}$ and the precision in the slow scan direction is 3.5 x 10$^{-3}$. The precision calculated here is for the spatial resolution of 2.8 nm. Improved probe stability in the instrument and faster response of the scan coils (reducing the recording time) could further improve this performance in the future. Rotating the scan direction by 90° provides a direct way to trade precision performance from the y to z axis at the expense of an extra image acquisition.

The typical trade-off between spatial resolution and precision in GPA is also present here as demonstrated in Figure.11 for different low pass filtering choices. The advantage is that this choice can be made during post processing [14] [15]. The typical mask size for low pass filtering option is much higher in the case of the proposed moiré quadrature demodulation as opposed to applying GPA directly on moiré images. The reliability of strain maps has significantly increased alongside the spatial resolution. The precision of strain measurement observed for different mask sizes in Figure.11a is $1.1 \times 10^{-3}$, $1.7 \times 10^{-3}$, $2.3 \times 10^{-3}$ for mask sizes of $1/5$, $1/3$ and $1/2$ (which leads to approximately ≈ 2.8 nm, 1.7 nm and 1.2 nm spatial resolution) times the sampling frequency. This once more demonstrates the inverse nature of precision vs spatial resolution for GPA. [15] [16]

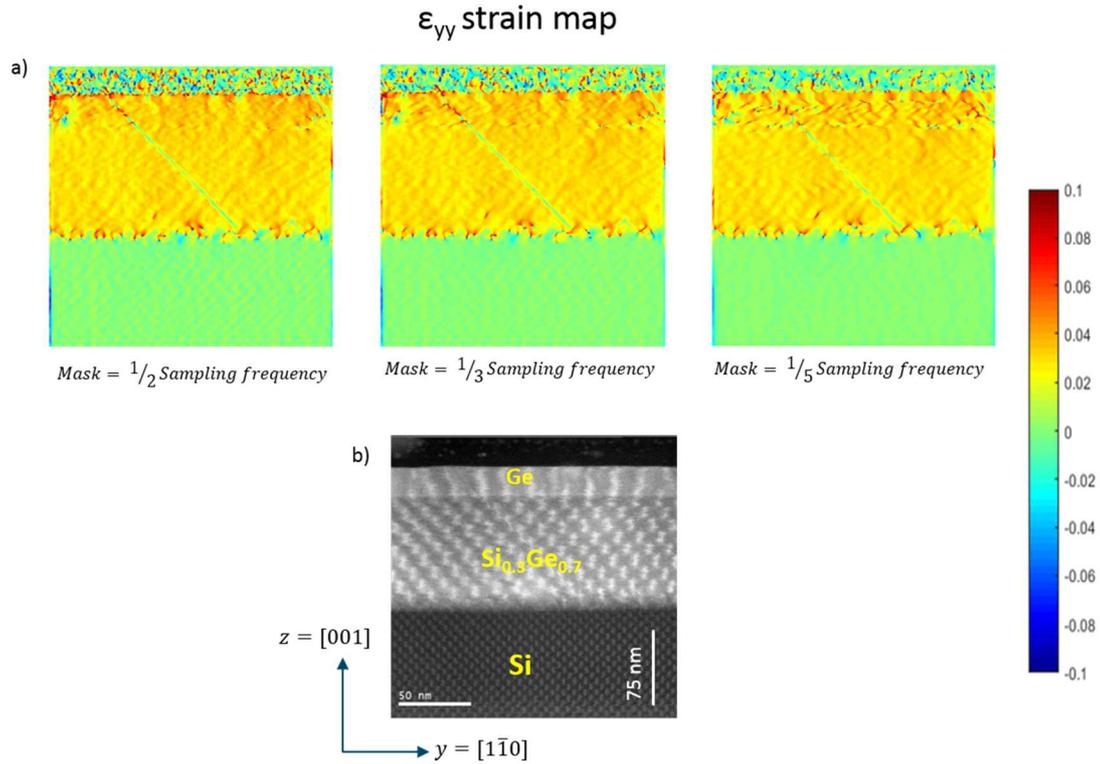

*Figure 11 a) $\varepsilon_{yy}$ strain maps obtained using GPA after performing the phase shifting and Quadrature demodulation. The maps are obtained for different mask diameters covering up to half of the sampling frequency b) HAADF STEM moiré image of the Si and Ge sample used for strain analysis.*

Analysing the line profiles of $\varepsilon_{xx}$ reveals that the Ge and $Si_{0.3}Ge_{0.7}$ lattice distances are not the same as is expected for a fully strained Ge layer. This means there is relaxation of the Ge layer in the x direction. However, Ge conforms with the lattice distances of $Si_{0.3}Ge_{0.7}$ in the y direction as seen in the $\varepsilon_{yy}$ line profile. Since the relaxed Ge lattice is larger than that of $Si_{0.3}Ge_{0.7}$, it means that the Ge is compressively strained in the y direction. This results in an elongation in the z direction due to the Poisson effect [17] and can be seen in the $\varepsilon_{zz}$ line profile. The strain values computed on the Ge channel region with respect to bulk Ge (Table 1) for the 16 nm FinFET reveals a uniaxial strain (y-axis) along the channel of the transistor. The strain is almost completely relaxed across (x-axis) the channel and there is a small tensile strain along the growth direction (z- axis). This trend is also confirmed after comparison with the diffraction based strain measurement techniques: NBD and Bessel diffraction [3] [4].

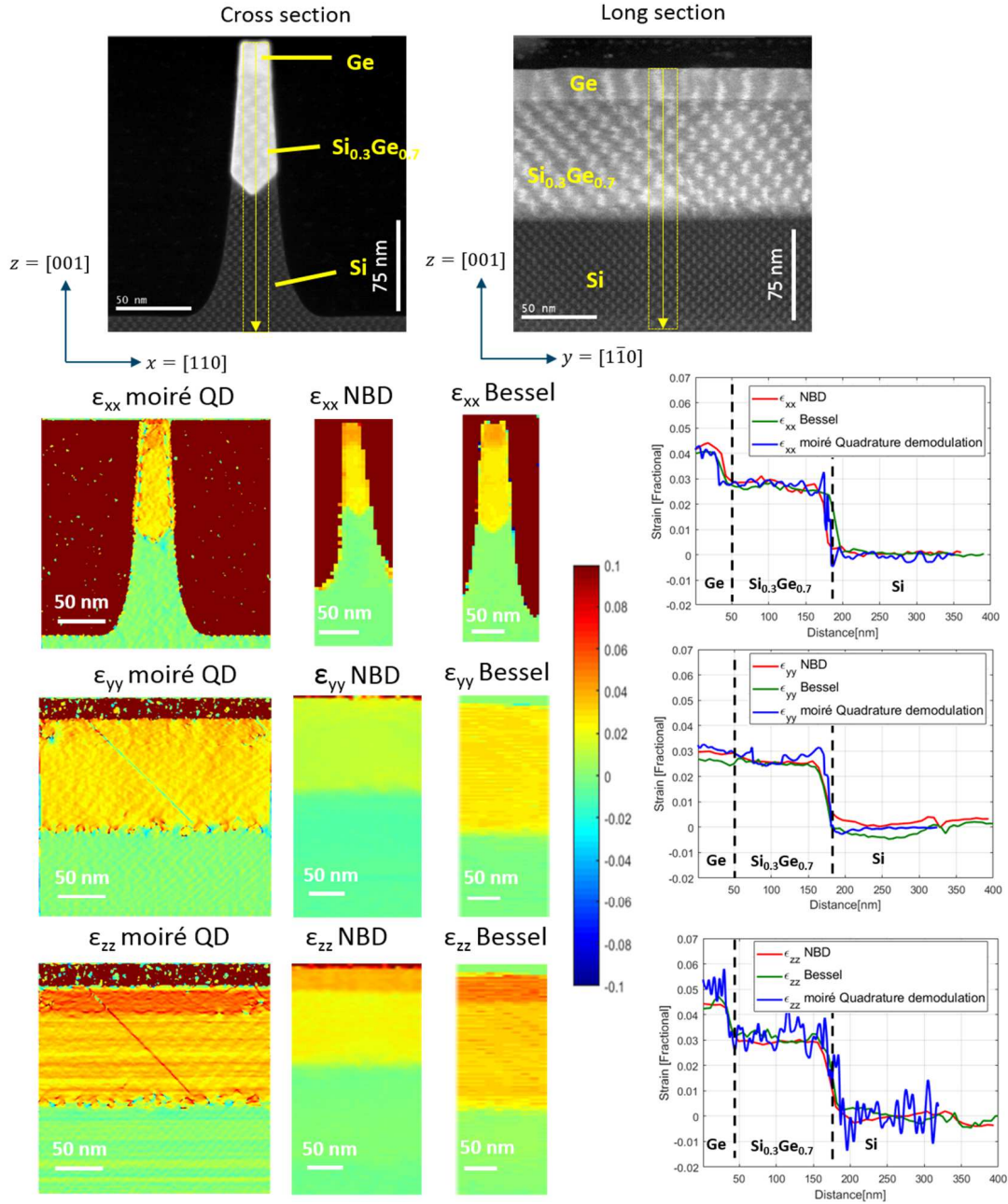

*Figure 12 Strain maps obtained using moire Quadrature demodulation(QD), Bessel and NBD for the 16nm FinFET on the cross section and long section of the Si and Ge samples providing an estimate of the complete 3D strain over the full structure. Strain profiles averaged over 16 nm horizontally as indicated by the yellow rectangle in the moiré images for all the three techniques mentioned above.*

Table 1 shows a comparison of the strain values in the channel, refered to bulk Ge for moiré quadrature demodulation with NBD and Bessel diffraction.

Table 1 Normal strain values in the germanium channel. The value is averaged over the germanium region and standard deviation is obtained for the complete Ge region.

| Technique | $\varepsilon_{xx}$ Ge (across the channel) (%) | $\varepsilon_{yy}$ Ge (along the channel) (%) | $\varepsilon_{zz}$ Ge (Growth direction) (%) |
|---|---|---|---|
| | **Standard deviation over the full Ge area** | | |
| Bessel diffraction | -0.17 ± 0.02 | -1.48 ± 0.02 | 0.18 ± 0.05 |
| Nano-beam diffraction | -0.20 ± 0.04 | -1.81 ± 0.03 | 0.18 ± 0.04 |
| Moiré quadrature demodulation | -0.26 ± 0.08 | -1.37 ± 0.03 | 0.12 ± 0.03 |

The strain values are averaged over the complete Ge region and the standard deviation represented here is over the full region and does not represent the deviation for each individual pixel in the image. Standard deviation over an area $= \frac{\text{Std.deviation}}{\sqrt{area}}$. This allows to make a fair comparison between methods which have a different spatial resolution and number of measurement points in the strain maps.

Table 2 Performance comparison of the three techniques evaluated in the strain free reference area: Bessel, NBD and moiré quadrature demodulation.

| Method | Spatial resolution | Precision over an area | Precision at nominal resolution | Field of view | Acquisition time (per pixel) | Analysis time (per pixel) | Total dose $[e^-/nm^2]$ |
|---|---|---|---|---|---|---|---|
| Bessel diffraction | 1-3 nm | $\varepsilon_{xx}$ = 4.2x10$^{-5}$ per 20x50 nm$^2$<br><br>$\varepsilon_{yy}$ = 4.4x10$^{-5}$ per 50x50 nm$^2$<br><br>$\varepsilon_{zz}$ = 9.2x10$^{-5}$ per 50x50 nm$^2$ | $7 \times 10^{-4}$ | > 500 x 500 nm$^2$ | ≈150 ms | ≈124 ms | 4.6x10$^6$ |
| Nano – beam diffraction | 1-3 nm | $\varepsilon_{xx}$ = 6.3x10$^{-5}$ per 20x50 nm$^2$<br><br>$\varepsilon_{yy}$ = 7.1x10$^{-5}$ per 50x50 nm$^2$<br><br>$\varepsilon_{zz}$ = 7.2x10$^{-5}$ per 50x50 nm$^2$ | $4 \times 10^{-4}$ | > 500 x 500 nm$^2$ | ≈ 25 ms | ≈156 ms | 0.7x10$^6$ |

| Moiré quadrature demodulation | ≥ 0.7 nm | $\varepsilon_{xx}$ = 3.1x10$^{-4}$ per 20x50 nm$^2$ $\varepsilon_{yy}$ = 2.6x10$^{-4}$ per 50x50 nm$^2$ $\varepsilon_{zz}$ = 2.5x10$^{-4}$ per 50x50 nm$^2$ | $1 \times 10^{-3}$ | < 500 x 500 nm$^2$ (Instrumental limitations apply due to scan coil response) | ≤ 0.1 ms | < 4 µs or real time possible | 0.1x10$^6$ |
|---|---|---|---|---|---|---|---|

Table 2 shows the comparison of the moiré Quadrature demodulation technique with both diffraction-based techniques. The acquisition time and the analysis time per pixel is significantly lower for the moiré method as opposed to diffraction techniques. This is a direct consequence of the low frame rate that is present in any CCD camera as compared to the very fast readout possible with the HAADF detector used in the moiré method. Even the latest camera technology is still an order of magnitude slower and will probably remain so for the foreseeable future.

The diffraction techniques involve 4D-STEM mechanism where each individual diffraction pattern is stored per pixel and hence it is also expensive in terms of data size in comparison to the moiré technique. Finally, The precision of the moiré Quadrature demodulation technique is $1 \times 10^{-3}$ computed for 3nm spatial resolution and is in close agreement with the high resolution STEM and standard STEM moiré techniques [18] [19] [20] but with substantial increase in spatial resolution compared to standard moiré and a higher field of view compared to high resolution STEM. The boosting factor that moiré applies on the strain has an added advantage of effectively detecting small amounts of strain.

Due to slower acquisition times, the total electron dose(number of electrons per area [$e^-/nm^2$]) impinged on the sample is significantly higher for the diffraction techniques. This could change in the future if pulsed sources or fast shutters are becoming available that would avoid illuminating the sample while waiting for the camera readout. Also it is to be noted that the higher the dose, the higher is the signal to noise ratio and this could have a direct impact on the precision of measurement. Hence, it is important to define precision relative to the square root of electron dose $\left(P_d = precision \times \sqrt{Dose} \left[\sqrt{e^-}/nm\right]\right)$. The $P_d$ for NBD = 1.8 $\left[\sqrt{e^-}/nm\right]$, for Bessel = 2.5 $\left[\sqrt{e^-}/nm\right]$ and for moiré Quadrature demodulation technique = 0.1 $\left[\sqrt{e^-}/nm\right]$. Hence, the moiré Quadrature demodulation technique is also highly dose efficient in comparison to the diffraction based strain measurement techniques.

In summary, the moiré quadrature demodulation technique offers some of the important practical advantages over diffraction techniques such as fast data acquisition, no need for specialised high speed diffraction camera's, very fast data analysis and flexible choice of spatial resolution.

## Conclusion

Standard strain mapping using GPA on moiré images suffers from reduced spatial resolution and interference from unwanted fringes and can lead to unreliable strain maps. A novel demodulation technique based on a generalisation of quadrature demodulation is proposed. We have shown that this significantly improves the spatial resolution and gives a more reliable and precise strain map at the expense of a somewhat longer acquisition as compared to conventional moiré imaging. The technique offers flexibility in the choice of field of view especially when compared to atomic resolution HR(S)TEM based methods. The method is near real-time in terms of acquisition and data analysis and provides the capability to trade spatial resolution for precision. Since moiré has the ability to boost strain values, it is highly sensitive and strain precision can be comparable to diffraction based methods without the need for specialised and slow diffraction cameras. We demonstrated the method on a relevant semiconductor device sample and compare its performance with diffraction based techniques.

## Acknowledgements

The Qu-Ant-EM microscope and the direct electron detector used in the diffraction experiments was partly funded by the Hercules fund from the Flemish Government. This project has received funding from the GOA project "Solarpaint" of the University of Antwerp. We would also like to thank Dr. Thomas Nuytten and Prof. Dr. Wilfried Vandervorst from IMEC, Leuven for their continuous support and collaboration with the project.

## Appendix A

The lattice spacing and the atomic positions in the real space image have direct relation with the sinusoidal components in the FFT. This is not true for a diffraction pattern, which, has an inverse relation. Since we are dealing with GPA and real space images, we can directly relate the frequency components with the atomic positions as normally done in GPA.

The reference frequency is represented as $g_{ref}$ and the strained frequency as $g_s = g_{ref}(1+\delta)$, $\delta$ is the strain. After sampling with the STEM probe with frequency $f_s$ satisfying the Nyquist criterion, the digital frequency of the reference and strained area becomes $\frac{g_{ref}}{f_s}$ and $\frac{g_s}{f_s}$. This can also be called the strain calculated for the high resolution images which always satisfy the Nyquist criterion.

$$Strain_H = \frac{g_s/f_s - g_{ref}/f_s}{g_{ref}/f_s} = \delta$$

So the obtained strain value is independent of the sampling frequency as long as the Nyquist criterion is satisfied.

Now, consider the moiré created using a different sampling frequency $f_m$ that do not satisfy the Nyquist criterion, so the digital frequency obtained now is

$$g_{ref,m} = \frac{g_{ref}}{f_m} - n$$

$$g_{s,m} = \frac{g_s}{f_m} - n = \frac{g_{ref}(1+\delta)}{f_m} - n$$

$n$ is the moiré window from which the frequency is translated (Figure.1). Now, the strain obtained by moiré is

$$Strain_M = \frac{\frac{g_s}{f_m} - n - \left(\frac{g_{ref}}{f_m} - n\right)}{\frac{g_{ref}}{f_m} - n} = \frac{\delta}{1 - \frac{nf_m}{g_{ref}}}$$

$$Strain_M = \alpha \, Strain_H$$

Where $\alpha = \frac{1}{1 - \frac{nf_m}{g_{ref}}}$ is the amplification factor.

So, the strain values for the moiré do depend on the sampling frequency and the moiré window from which the frequency is translated